# Post-processing of Engineering Analysis Results for Visualization in VR System

Stoyan Maleshkov[1], Dimo Chotrov[1]

[1] Virtual Reality Lab, Technical University Sofia, Sofia, 1000, Bulgaria

**Abstract**
The applicability of Virtual Reality for evaluating engineering analysis results is beginning to receive increased appreciation in the last years. The problem many engineers are still facing is how to import their model together with the analysis results in a virtual reality environment for exploration and results validation. In this paper we propose an algorithm for transforming model data and results from finite element analysis (FEA) solving application to a format easily interpretable by a virtual reality application. The algorithm includes also steps for reducing the face-count of the resulting mesh by eliminating faces from the inner part of the model in the cases when only the surfaces of the model is analyzed. We also describe a possibility for simultaneously assessing multiple analysis results relying on multimodal results presentation by stimulating different senses of the operator.
***Keywords:*** *Virtual Reality, Finite Element Analysis, Engineering Analysis Results Validation, CAD to VR Export, Multimodal Presentation.*

## 1. Introduction

The benefits of using virtual reality (VR) for the evaluation and simulation of CAD models and their behavior and its application in the engineering field have been long realized. Some such applications together with main problems and challenges lying before integrating CAD and VR have been discussed in [3].

In the last years VR has moved from just a fancy visualization tool to an important means for improving and speeding up the solution of engineering tasks. In [4] the authors describe their software system VtCrash which allows not only immersive observation of a pre-calculated crash simulation but also manipulation of the observed crash. [1] discusses some application possibilities of VR in the field of chemical engineering and points out some advantages it gives for student education. The authors have developed different algorithms allowing for faster user interaction with complex scenes. An interesting method for combining VR and CAD by incorporating VR techniques directly in a CAD application is described in [2]. The authors use the development application programming interface (API) of the CAD application to perform the calculation and visualization of scene illumination in the CAD application itself, allowing the user to export the scene and results to a VRML file.

Although the applications of VR for engineering tasks are improving rapidly the possibilities it offers are far from a satisfactory realization and give space for a lot of research. Still the term most often connected with VR is visualization meaning only visual presentation. There are still few scientific researches in the field of using other human senses for data presentation. One such example is the research discussed in [7] dealing with performing acoustic simulation for analyzing noise behavior.

Based on our previous research experience in the field of multimodal presentation of object features described in [6] we present in this paper an algorithm for post processing results from different engineering analyses (focused on finite element analysis (FEA)) that allows their simultaneous presentation in a VR environment by stimulating multiple senses of the user.

## 2. Convert FEA results to a form suitable for VR presentation

The Finite Element Analysis (FEA) is usually performed to simulate the behavior of some part or fluid under specific conditions. The analysis often includes as an essential step importing and tessellating a CAD model to perform the necessary calculations. The results obtained from the FEA solver are for the new geometry after the tessellation. This makes the CAD model geometry unusable when one wants to present the results in VR. This means that the mesh of the model used for the analysis has to be exported together with the calculated results so that they can be interactively and adequately presented in a VR environment.

## 2.1 Finite element analysis background

The Finite Element Method (FEM) can be used to solve tasks in engineering areas like structural analysis, heat transfer, fluid flow and electromagnetic potential. It is applied when no analytical solution can be obtained or the solution is too complex to be calculated for an examined geometrical model. The FE method involves representing the geometry as a system of smaller and simpler bodies (finite elements), a process called discretization or tessellation (see Fig. 1). Then the equations are solved for each of the finite elements and the results are combined to receive an approximate solution [5].

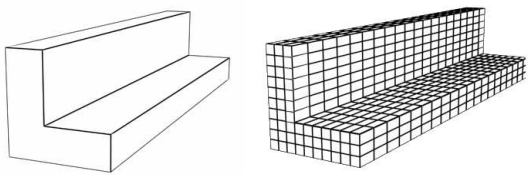

Fig. 1. The discretization process

The discretization of the geometry and the results calculation are usually performed with an application often called *solver*. Examples for such applications are Ansys, Abaqus, Nastran. For the purposes of the research and experiments described in this article we have used the Ansys software package.

## 2.2 Transforming the mesh

The problem with the tessellation performed by the solver application is that the finite mesh and the calculation results are available during the analysis or can be saved in a native for the solver application format, which means they cannot be exported to some standard intermediate format supported by another application. The only way of obtaining information about them in Ansys is by exporting lists describing the nodes and elements of the mesh and their results values in text files.

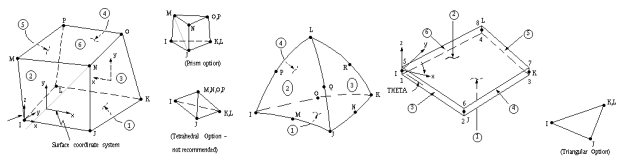

Fig. 2. Examples for element types in Ansys [9]

To describe the geometry of the model two lists are needed – a list of the elements building the mesh and a list of the nodes building the elements. The format of the node list is simple – it contains a sequential number of the node and its coordinates in space. The format of the element list is more complicated:

```
EL MAT TYP REL ESY SEC    NODES
21  1   2   1   0   21   23  24  65  65
22  1   2   1   0   22   22  23  60  60
23  1   2   1   0   23   80  26  27  27
24  1   2   1   0   24   64  22  60  60
```

The important columns in the list are the *NODES* column – which nodes build the element, and the *TYP* column – the type of the element or how the nodes are combined to build the element, which may often have an irregular form (see Fig. 2).

```
public ref class AnsysElement
{
public:
    AnsysElement()
    {
        nodes = gcnew List<int>();
        elementType = nodeMap = 0;
    }

    //node numbers, forming the element
    //contains -1 for inner nodes after they are
    List<int>^ nodes;
    //the count of the actual nodes in the list,
    int realNodeCount;
    //Types: 1 - box element; 2 - surface element
    int elementType;
    //shows which of the 8 nodes are present, aft
    //flag 1 means node is not to be used
    int nodeMap;
};
```

Code 1. Class describing an Ansys element

We transform the data from the nodes and elements lists to receive a new format which can easily be interpreted by a visualization application – the format describes vertices with their coordinates in space and triangles (polygons), formed by the vertices, which compose the geometry of the model. The vertices are directly mapped from the nodes of the finite model and their coordinates. The triangles are formed by splitting the elements. Each element is described by an instance of a special class *AnsysElement* (see Code 1), containing a list with the node numbers building the element, the type of the element, which determines how the element is triangulated, and a node map used for an optimization step, described later in this chapter.

```
List<List<int>^>^ lstTriPolygons;
switch(dicElementType[kvp.Value->elementType]) //create polygons dependi
{
case AnsysElementType::AET_Shell: //shell elements can contain 3 or 4 ve
    {
        if(kvp.Value->realNodeCount == 3)
        { //already a triangle, only remove -1 entries
            lstTriPolygons = gcnew List<List<int>^>();
            lstTriPolygons->Add(gcnew List<int>());
            for(int i = 0; i < kvp.Value->nodes->Count; i++)
                if(kvp.Value->nodes[i] != -1)
                    lstTriPolygons[0]->Add(kvp.Value->nodes[i]);
        }else if(kvp.Value->realNodeCount == 4)
        { //a quad use CreateTriangles from quad directly, after removin
            lstTriPolygons = gcnew List<List<int>^>();
            lstTriPolygons->Add(gcnew List<int>());
            for(int i = 0; i < kvp.Value->nodes->Count; i++)
                if(kvp.Value->nodes[i] != -1)
                    lstTriPolygons[0]->Add(kvp.Value->nodes[i]);
            lstTriPolygons = CreateTrianglesFromQuad(lstTriPolygons[0]);
        }
    }
    break;
case AnsysElementType::AET_Solid92:
    {
        lstTriPolygons = TriangulateSolid92Element(kvp.Value);
    }
    break;
```

Code 2. Element triangulation

The algorithm passes through each element of the Ansys model and divides it into triangles depending on its type – see Code 2. The type of the element is crucial because it describes how the nodes of the element are interconnected and hence – how the element itself can be divided into triangles. After all elements have been processed the result is a list of triangles building the model.

The full workflow for transforming the ANSYS exported nodes and elements lists to a format easily interpretable by a VR application is shown on Fig. 3.

For complex geometry models the high number of vertices and faces could lead to slower visualization and response-time by the VR application. To reduce this impact we include an option for optimizing the number of polygons of the model in the case when only the surface of the model will be of interest during the VR presentation. The optimization reduces the number of nodes and triangles by excluding element nodes lying in the inner part of the model and building polygons only from element sides lying on the surface of the model. For the purpose two lists of nodes have to be exported from the solver application and specified as input for the conversion process – one list containing all the nodes and a second one containing only nodes lying on the surface of the model. During the conversion process a node map is generated for each element, based on the intersection of the full and partial node list.

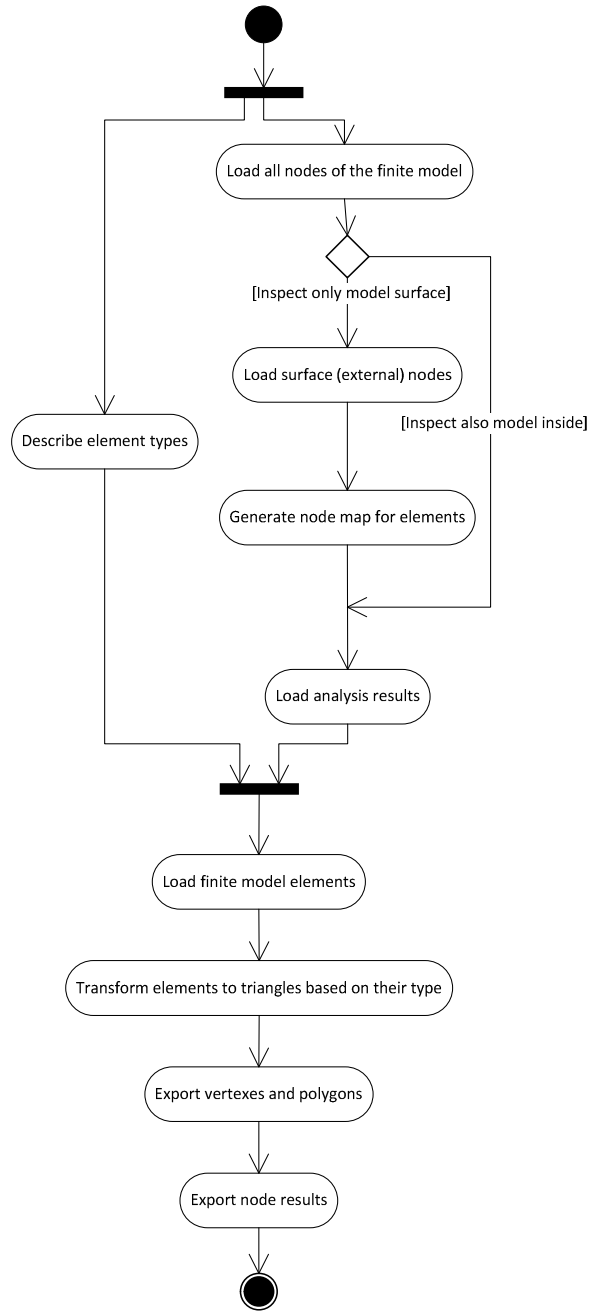

Fig. 3. Ansys to VR data transformation

The node map links a sequence of bits to each node: '1'-s for the element nodes that should be skipped and '0'-s for the nodes of the element that should remain. The resulting triangles in the output are built depending on the node map value. As some of the original nodes are excluded from the output the resulting vertices of the model have to be renumbered.

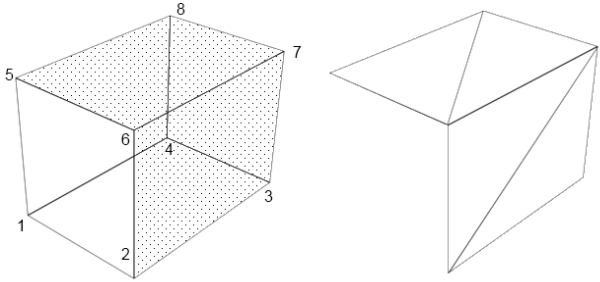

Fig. 4. Mesh optimization

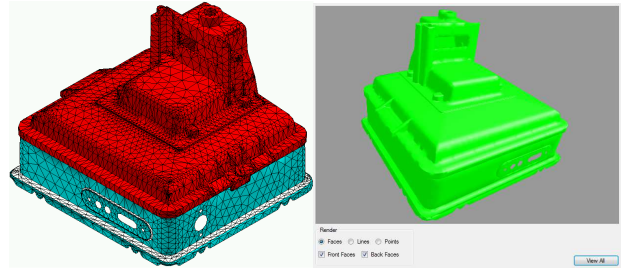

Fig. 5. Ansys model and the resulting VR model

Fig. 4 shows an example with a simple parallelepiped element which lies on an edge of a model. Nodes with numbers 2, 3, 5, 6, 7 and 8 are lying on the surface, while nodes 1 and 4 are inner nodes (Fig. 4 – left). Respectively the node map for this element would be 00001001 giving a value of 9 – the optimization skips nodes 1 and 4 and builds triangles only from the remaining surfaces thus reducing the node and polygon count of the resulting model (Fig. 4 – right).

The result lists exported from Ansys are simply node numbers with their result values so no special steps are needed for the result file transformation. Only if some nodes have been excluded, their results have also to be omitted in the results file generated by the transformation process.

## 3. Exploring and Evaluating Finite Element Analysis results in VR

For the exploration of the model and engineering analysis results in VR environment we use our own custom developed VR software system, described in [8]. The results from the engineering analysis are treated as implicit object properties that can be presented to the user through different modalities.

### 3.1 Importing the mesh in the VR application

After the finite element model has been processed as described earlier loading the mesh of the model in the VR application is quite simple – read the vertices and their coordinates from the file and create the polygons of the mesh as specified by the triangles description in the file. After that for each polygon its normals have to be calculated. Fig. 5 shows an original Ansys finite element model (left) and its equivalent model loaded in the VR application.

### 3.2 Assigning results data to the mesh vertices

After the model has been imported in the VR application it has to be associated with the results from the engineering analysis that are going to be presented and evaluated. For this purpose the converted results files exported from Ansys are used. Each results file contains the values for the vertices of the model from a specific analysis that has been performed. When opening a results file for the current model each value is associated with its vertex. Each vertex can have several results values associated with it.

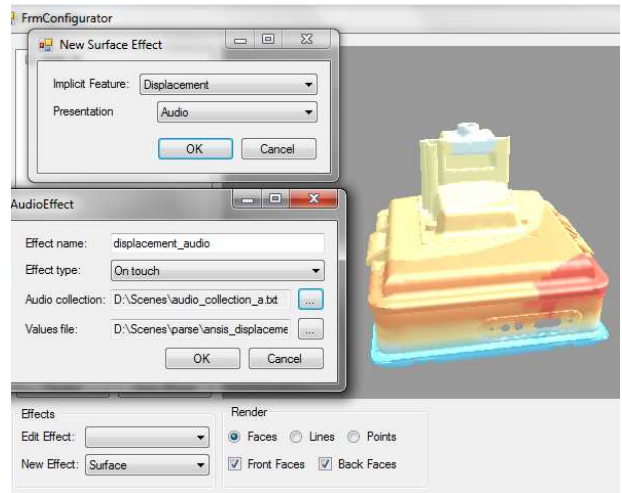

Fig. 6. Specifying analysis results and type of effects for presentation

Our approach allows the use of different sensorial stimuli, which we call effects, for the presentation of the analysis results. The user has to specify how the analysis results are going to be presented – visually, through audio or tactile stimulation. The use of different modalities for the results presentation allows the simultaneous evaluation of results from different analyses. Fig. 6 shows the assignment of displacement analysis results to the model of a boiler that are going to be presented to the user through the audio channel. The surface mesh representing the boiler geometry has already been assigned temperature analysis results which are being displayed visually through color codes.

### 3.3 Exploring and evaluating the model in VR

After the results from all analyses that are going to be explored and evaluated have been assigned to the surface mesh representing the object in the VR application the model is ready for presentation. The user can observe the model or scene in stereo and receive information about analysis results depending on the type of presentation effects that have been chosen.

As a test case we have performed temperature and displacement FE analyses of a boiler. In our VR application we have assigned visual effect for the presentation of the temperature distribution results and audio effect for the displacement distribution results. In this way we can observe directly the temperature distribution for the model based on its coloring and simultaneously receive audio feedback about the displacement at a specific place by pointing or clicking at the desired area of the object– see Fig. 7.

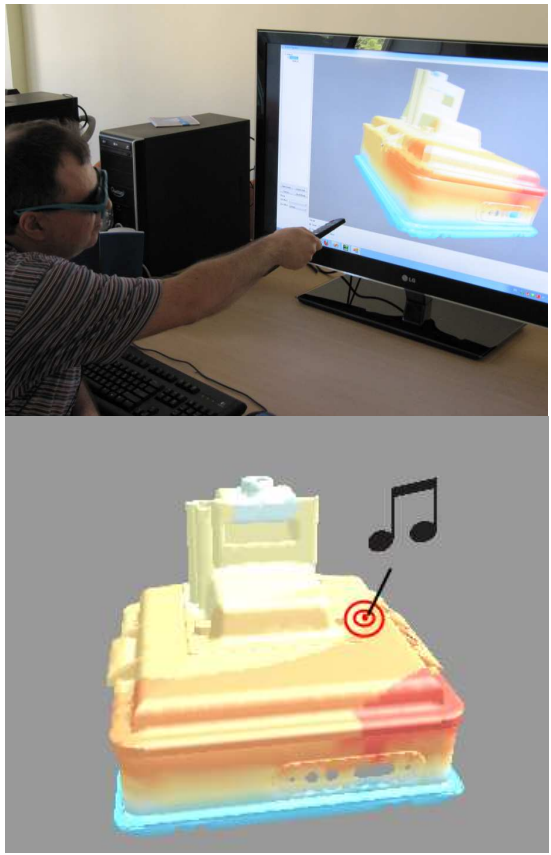

Fig. 7. Evaluating analysis results

### 4. Conclusion

Our solution allows the presentation of engineering analysis, specifically finite element analysis, results in virtual reality environment. For the purpose we export the finite element model geometry and results from the solver application and convert it to a form easily interpretable by a VR application. The results can be presented not only visually but also by stimulating other senses of the person evaluating the results. Our approach significantly improves user perception and understanding of the results – especially when evaluating multiple analysis results simultaneously. This is due to the application of multimodal presentation which allows for delivering more information to the user by distributing it among different perception channels. Transmitting information by stimulating different senses leads to faster, better and more intuitive evaluation of the results enhances the user interaction with the system and increases the sense of immersion.

### Acknowledgments

The authors wish to thank for the support of the National Science Found at the Bulgarian Ministry of Education, Youth and Science received through grant DDBY02/67-2010.

**Stoyan Maleshkov** has Eng. degree in system and control engineering (1975), master in applied mathematics (1977) and PhD in computer aided system design (1981), all received from the Technical University (TU) of Sofia, Bulgaria. Fulbright scholar (1989–1990) at Interactive Modeling Research Lab, Louisiana State University, Baton Rouge, USA. Associate Professor of computer aided engineering and computer graphics (1990) at TU Sofia. Department chair (2000-2004) and vice dean (2004-2008), both at the TU Sofia. Since 2008: Head of the Virtual reality lab, TU Sofia. Associate professor of computer graphics at the New Bulgarian University Sofia, as a second job (2000). IEEE member.

**Dimo Chotrov** has received BSc. (2007) and MSc. (2009) degrees in computer systems and technologies from the Technical University of Sofia. Currently he is PhD student, acting at the Virtual reality lab. Member of the IEEE.